\documentclass[12pt,a4paper]{article}
\usepackage[utf8]{inputenc}
\usepackage{amsmath}
\usepackage{amsmath}
\usepackage{mathtools}
\usepackage{amsfonts}
\usepackage{amssymb}
\usepackage{physics}
\usepackage{ dsfont }
\usepackage{braket}
\usepackage{siunitx}
\usepackage{graphicx}
\usepackage{enumerate}
\usepackage{sidecap}
\usepackage[a4paper, total={6in, 8in}]{geometry}
\usepackage [autostyle, english = american]{csquotes}
\MakeOuterQuote{"}
\usepackage{float}
\usepackage[toc,page,title]{appendix}
\usepackage{multirow}
\usepackage{amsthm}
\usepackage{mathtools}
\usepackage{braket}
\usepackage{sidecap}
\usepackage{eucal}
\usepackage{comment}
\usepackage{authblk} 
\usepackage{enumerate}
\usepackage{setspace}
\usepackage[super]{nth}
\usepackage{hyperref}
\usepackage{todonotes}
\usepackage{amsthm}
\usepackage{dirtytalk}

\usepackage{tikz}
\usetikzlibrary{backgrounds}

\usepackage[capitalise]{cleveref}
\Crefname{lemma}{Lemma}{Lemmas}
\Crefname{proposition}{Proposition}{Propositions}
\Crefname{definition}{Definition}{Definitions}
\Crefname{theorem}{Theorem}{Theorems}
\Crefname{conjecture}{Conjecture}{Conjectures}
\Crefname{corollary}{Corollary}{Corollaries}
\Crefname{example}{Example}{Examples}
\Crefname{section}{Section}{Sections}
\Crefname{appendix}{Appendix}{Appendices}
\Crefname{figure}{Fig.}{Figs.}
\Crefname{equation}{Eq.}{Eqs.}
\Crefname{table}{Table}{Tables}
\Crefname{item}{Property}{Properties}
\Crefname{remark}{Remark}{Remarks}
\Crefname{axioms}{Axioms}{Axioms}

\crefformat{footnote}{#2\footnotemark[#1]#3}

\makeatletter
\newtheorem*{rep@theorem}{\rep@title}
\newcommand{\newreptheorem}[2]{\newenvironment{rep#1}[1]{ \def\rep@title{#2 \ref{##1}} \begin{rep@theorem}}{\end{rep@theorem}}}
\makeatother

\newtheorem{thm}{Theorem}
\newreptheorem{thm}{Theorem}
\newtheorem{lemma}[thm]{Lemma}
\newreptheorem{lemma}{Lemma}
\newtheorem{col}[thm]{Corollary}
\newreptheorem{col}{Corollary}
\newtheorem{defn}[thm]{Definition}

\newtheorem{proposition}[thm]{Proposition}

\definecolor{ma}{rgb}{246,76,246}

\definecolor{ha}{RGB}{246,76,246}
\definecolor{ma}{rgb}{0.858, 0.188, 0.478}

\title{Encodings of Observable Subalgebras}

\author[1]{Maite Arcos\thanks{\texttt{maite.arcos@ucl.ac.uk}}}
\author[2]{Harriet Apel\thanks{\texttt{harriet.apel.19@ucl.ac.uk}}}
\author[2,3]{Toby Cubitt\thanks{\texttt{t.cubitt@ucl.ac.uk}}}

\affil[1]{Department of Physics and Astronomy, University College London, UK}
\affil[2]{Department of Computer Science, University College London, UK}
\affil[3]{Phasecraft Ltd.}

\date{March 2025}

\usepackage[backend=bibtex,style=alphabetic]{biblatex}
\begin{document}

\maketitle

\begin{abstract}
Simulating complex systems remains an ongoing challenge for classical computers, while being recognised as a task where a quantum computer has a natural advantage.
In both digital and analogue quantum simulations the system description is first mapped onto qubits or the physical system of the analogue simulator by an encoding.
Previously mathematical definitions and characterisations of encodings have focused on preserving the full physics of the system.
In this work, we consider encodings that only preserve a subset of the observables of the system.
We motivate that such encodings are best described as maps between formally real Jordan algebras describing the subset of observables.
Our characterisation of encodings is general, but notably holds for maps between finite-dimensional and semisimple $C^{*}$-algebras.
Fermionic encodings are a pertinent example where a mathematical characterisation was absent.
Our work applies to encodings of the the full CAR algebra, but also to encodings of just the even parity sector, corresponding to the physically relevant fermionic operators.
\end{abstract}


\section{Introduction}

A better understanding of complex phenomena in quantum systems is one of the motivating goals of quantum technologies and has led Hamiltonian simulation to attract much attention since the original conception \cite{Feynmansimulation}.
Part of the attraction is the wide-reaching effects of improving simulations in medicine \cite{proteinfolding,medicine}, chemistry \cite{ciracsimulation}, and even climate science \cite{preskillquantumcomp40,simulationreview}.
In contrast to digital simulation, analogue simulation involves directly engineering a simulator system to emulate a target system.
With theoretical evidence suggesting that analogue simulation does not require error correction \cite{Cubitt2019} or scalability \cite{ciracsimulation}, it is believed to be more accessible to near-term hardware.
This form of simulation has already proved effective in probing properties of lattice and spin models experimentally, allowing physicists to study features of simple quantum systems that were previously inaccessible \cite{Porras,Jaksch2004,Peng2010,localizationtransin2D}.
There is hope that when larger, more stable quantum processors are available, digital simulation will have a greater impact on industry.
There is a need for better theoretical descriptions of the encodings required to either map the system of interest onto the native operations of the analogue simulator or onto the qubits of a gate-based quantum computer.

The question of what it means mathematically for one system to encode the physics of another has been investigated in many-body physics for certain strong notions of simulation \cite{Bravyi2014,Cubitt2019,Apel2023}.
One common weakness of these works, as noted in \cite{Harley2023}, is that the notion of simulation considered stipulates the entire physics of the target system to be reproduced in the low-energy subspace of the simulator, precluding the possibility of simulating only part of the target system.
The characterisations in \cite{Cubitt2019,Apel2023} essentially prove a common mathematical form for encoding maps, whereby if the full physics is preserved, the only possible simulations consist of taking copies of the target system or its complex conjugate and applying a global isometry.
This raises the question of how the mathematical form varies if only a subset of observables are required to be simulated and whether this relaxation allows for new types of encoding mappings.

Understanding how only part of the target physics of a system can be reproduced in another system is a natural relaxation, with applications to both digital and analogue simulation, as well as verification protocols \cite{verificationsimulation}.
An example is the simulation of spin-lattice systems, where the initial Hamiltonian is described in terms of fermionic creation and annihilation operators and requires a fermionic encoding \cite{NielsenJordanWigner, BravyiKitaev}.
Here, only fermionic operators need to be preserved by the map, providing a motivating example where enforcing the entire physics to be replicated is superfluous.

This work formalises the notion of encoding a subset of observables.
We will argue that defining a notion of simulation for only part of the physics requires shifting the focus to an algebraic description, in agreement with what has been found in the high-energy physics literature \cite{harlowsubalgebras,targetspaceEEmanzenc}.
The main contribution is to connect well-established results concerning algebraic maps to the context of simulation via identifying the physically motivated subset of observables and how the ideas of encoding translate to operational requirements.

The rest of this paper is structured as follows.
After introducing some notation, \cref{sect def sim} begins by defining the minimal axioms of an encoding, which leads to the identification of the relevant set of observables as a subalgebra of a finite-dimensional JB algebra.
\cref{sect characterise} proves a full mathematical characterisation for the encodings defined for both simple algebras and semisimple algebras—where subalgebras of observables exhibit additional structure, such as superselection sectors.
We find that while for simple algebras the result agrees with that of \cite{Cubitt2019}, when the algebra of observables simulated has superselection rules, there is a slight relaxation on the mathematical form.
We close with a discussion of these differences and how the algebra of fermions fits into our framework as an example.

\subsection*{Notation and terminology}

In the following, $M_{n}(\mathbb{F})$ will denote the set of $(n \times n($ matrices over a field $\mathbb{F}$.
We will denote the set of $(n\times n)$ hermitian matrices as $\mathrm{Herm}_{n}(\mathbb{F})$, and write $a^{\dagger}$ to denote the hermitian conjugate of the element $a$.

For any ring $R$ of characteristic not two, we will write the (associative) multiplication of two elements $a,b \in R$ as $ab$, and will denote the additive and multiplicative identities as $0_{R}$  and   $\mathds{1}_{R}$ respectively. If $R$, $R'$ are rings, we will say that a mapping $\Gamma: R \rightarrow R'$ is a ring homomorphism if $\Gamma$ is a homomorphism of the underlying additive groups which satisfies $\Gamma(ab)=\Gamma(a)\Gamma(b)$ and $\Gamma(\mathds{1}_{R})=\mathds{1}_{R'}$.

An associative algebra over a commutative ring \( R \) is a ring \( A \) which is also a module over \( R \), such that the ring and module multiplication are compatible in the following way: $x(ab) = (xa)b = a(xb) \quad \text{for all } x \in R, a, b \in A$. \( A \) is also called a \( R \)-algebra. When \( R \) is a field, a basis for \( A \) as a module over \( R \) is said to be a basis for the algebra \( A \), and \( A \) is said to be a finite-dimensional \( R \)-algebra, if \( A \) is finite dimensional as a module over \( R \). A subset $S$ of an algebra $A$ is called a subalgebra if it is a submodule of $A$ which is also an algebra. A finite-dimensional algebra is semisimple if it is isomorphic to a finite direct sum of simple algebras, or, equivalently, if its Jacobson radical is zero.

A Jordan algebra $A$ is a non-associative algebra over a field with a bilinear mapping $A \times A \rightarrow A$ $(a, a') \mapsto a \circ a'$ which is commutative and satisfies the Jordan identity $ a \circ (b \circ a^{2}) = (a \circ b) \circ a^{2}$.
Given an associative algebra of characteristic not two over a field $\mathbb{F}$, a special Jordan algebra is a linear subspace in which the Jordan product corresponds to $a \circ b = \frac{1}{2}(ab+ba)$.
 Given two Jordan algebras $A$ and $B$, a Jordan homomorphism is a linear mapping $\gamma: A \rightarrow B$  satisfying $\gamma(a^{2})= \gamma(a)^{2}$ and $\gamma(aba)= \gamma(a)\gamma(b)\gamma(a)$ or, equivalently for special Jordan algebras over a field of characteristic not two, $\gamma(ab+ba)= \gamma(a)\gamma(b)+ \gamma(b)\gamma(a)$.
A Jordan algebra $A$ is said to be formally real if, given $a_{i} \in A$, $\sum_{i}a_{i}^{2}=0 \implies a_{1}= ...=a_{n}=0$. A Jordan algebra $A$ is a Jordan Banach algebra if it is a Jordan algebra and it has a norm $\norm{\cdot}: A \rightarrow \mathbb{F}$ which satisfies $\norm{a \circ b} \leq \norm{a} \norm{b} \forall a,b \in A$.

Given a Jordan algebra $A$, two orthogonal idempotents $p,q$ are said to be strongly connected if there exists $v \in pAq$ such that $v^{2}= p + q$.
An idempotent $p$ is said to be minimal if, for any non-zero idempotent $q$, $pq=qp=q$ implies $p=q$.

We will use $\mathcal{A}$, $\mathcal{B}$ to denote associative algebras. We will say that an associative algebra $\mathcal{A}$ is an algebra with involution $*$ if there exists a mapping $*: \mathcal{A} \rightarrow \mathcal{A}$ $a \mapsto a^{*}$ satisfying $a^{**}= a$, $(a+b)^{*} = a^{*}+ b^{*}$ and $(ab)^{*}=b^{*}a^{*}$. We will use $A$,$B$ to denote the special Jordan algebras of elements satisfying $a^{*}=a$.
Given an element $a$ of a unital Banach algebra over a field $\mathbb{F}$, $\mathrm{spec}(a)$ denotes the set $   \lambda \in \mathbb{F} $ such that$ : a-\lambda \mathds{1}$ is not invertible.

 Let $M_{1}$ and $M_{2}$ be $R$-modules, and define their Cartesian product $M_{1} \times M_{2} = \{ (m_{1},m_{2}) : m_{1} \in M_{1} , m_{2} \in M_{2}\}$. Since we will be working with finite dimensional vector spaces only, we will write the Cartesian product of modules as the direct sum $\oplus$. If $M,N$ are $R$-modules, a mapping $f: M \rightarrow N$ is an $R$-module homomorphism if $f(m+m')=f(m)+f(m')$ and $f(rm)=rf(m)$ $\forall r \in R$ and $\forall m,m' \in M$.
 We will write $\mathrm{Hom}(M,N)$ to denote the set of $R$-module homomorphisms from $M$ into $N$.
 We will focus on the cases where $R=\mathbb{R}$ or $R= \mathbb{C}$. A module $M$ is simple if it contains no proper non-zero submodule.
 A module $M$ is semisimple if it is the direct sum of simple modules. A ring $R$ is simple if it is simple as an $R$-module. Recall that if a ring is simple, then there exists a division ring $D$ and a positive
integer n such that $R \cong M_{n}(D)$.

\section{Strong simulation of subsets of observables}\label{sect def sim}

This work considers simulation where only a subset (more precisely, a subalgebra) of observables on the target system is reproduced in the simulator system.
This framework is less restrictive than demanding that the entire physics is reproduced as in \cite{Cubitt2019, Apel2023}.
Moreover, in practical simulation applications it is very often the case that only certain observables of the system are of interest.

We demand a strong form of simulation where all measurement outcomes of the encoded observables are preserved.
This implies that the set of eigenvalues for an observable and its simulated counterpart agree, while degeneracies and eigenvectors are free to differ.
This condition is sufficient to ensure that all measurement outcomes for a target observable can be reproduced in the simulator system.
Taking the Hamiltonian itself to be part of the set of simulated observables of interest is sufficient to reproduce all thermal properties and time dynamics of the target in the simulated system \cite{Cubitt2019}. 
Spectrum-preserving maps between different algebraic structures have been studied in the mathematics literature, as well as connections to high-energy physics \cite{witten, spectrumpreserving}.

In addition to being spectrum-preserving, we require the encoding to be convex.
Operationally, a convex combination of observables corresponds to selecting an observable at random from the ensemble of observables according to some probability distribution, measuring the observable, and reporting the outcome.
Constraining the encoding to be convex is motivated by assuming that taking a probabilistic combination of simulated observables and considering the probabilistic combination as a new observable is indistinguishable.
Hence, \textbf{we will formally require that an encoding of a subset of observables be described by a map that is convex and spectrum-preserving}.

The domain and range of the encoding should not be the full algebra of observables, but a subset of observables.
This leads to the question: \emph{"What is a valid subset of observables?"} given that the simulator system is constrained to obey the usual rules of quantum mechanics.
Note, for example, that if observables, $a_1$ and $a_2$ are both simulated by a system, any probabilistic combination must also be simulated, which immediately shows that the simulated set is larger than just $\{a_1,a_2\}$.
Defining the exact set will lead us to the notion of the algebra of a subset of observables.

Note that for any observable $a$, scaling the observable by a real number $\lambda \in \mathbb{R}$ operationally corresponds to scaling the measurement outcome.
Therefore, the observable $\lambda a$ must be preserved by a simulation whenever $a$ is.
Real scaling with convexity implies that any sum of simulated observables is also preserved by the encoding.
A crucial example of an observable that is the natural sum of two other observables is the Hamiltonian $H= \lambda p^{2} + \mu x^{n}$, where $p$ and $q$ are the momentum and position operators, respectively, and $\lambda$, $\mu$ are real scalars.

By a similar argument to rescaling, one sees that the observable $a^{m}$ may be interpreted as the observable associated with taking the $m$'th power of the results obtained after measuring $a$.
This leads to (associative) powers of simulated observables also being in the simulated set.
Finally, combining powers with sums leads to requiring that the Jordan product $a_1 \circ a_2 = a_1 a_2 + a_2 a_1$ is also accessible in the simulator system since $(a_1 + a_2)^2 = a_1^2 + a_2^2 + a_1\circ a_2$.

We therefore define the algebra generated by a single observable as all real-linear combinations and Jordan products of the observable, in agreement with the literature \cite{Strocchireview, F.Strocchi}. 
Similarly, for multiple observables, we consider the set of all real-linear combinations and Jordan products of those observables.
It is easy to see that this forms a Jordan algebra (see \cref{appen C*} for more details) which is a Jordan subalgebra of the full algebra of observables.
Associating to each observable the number $ \norm{A} = \sup_{\omega} |\omega(A)| $ turns the considered set into a Jordan Banach algebra.
If we further restrict the algebras to be formally real, we arrive (for finite-dimensional systems) at the definition of a special JB algebra.
Such algebras are guaranteed to be simple or semisimple (see \cref{JordanvnWigner}).
Note also that there is only a nontrivial subalgebra of observables if the full algebra is semisimple, i.e., if it has superselection sectors \cite{jordanVNWigner}.

\begin{defn}[Special JB Algebra]\label{specialJBdefn} \footnote{ Here we are using the fact that, for finite dimensional algebras, a special Jordan algebra is a JB algebra if and only if it is formally real (see \cref{appen C*} and \cite{jordanoalgebrasHancheStormer} for technical details)} A special JB algebra is a special real Jordan algebra $A$ equipped with a complete norm, $\forall$ $a,b \in A$ satisfying 
\begin{enumerate}
    \item $\norm{a \circ b} \leq \norm{a} \norm{b}$;
    \item $\norm{a^{2}}=\norm{a}^{2}$;
    \item $\norm{a^{2}} \leq \norm{a^{2}+b^{2}}$.
\end{enumerate}
\end{defn}

We have motivated that the maps of interest are spectrum-preserving convex maps between subalgebras of observables, which leads to the following formal definition of an encoding:

\begin{defn}[Encoding]\label{defn sim map}
    Let $A$ and $B$ be two finite-dimensional special JB algebras.
    A map $\gamma: A \rightarrow B$ is an \emph{encoding} if it is spectrum preserving and convex in the following sense:
    \begin{enumerate}[(i.)]
        \item $\textup{spec}[\gamma(a)] = \textup{spec}[a]$
        \item $\gamma(\lambda a_1 + (1-\lambda) a_2) = \lambda \gamma(a_1) +(1-\lambda)\gamma(a_2)$
    \end{enumerate}
    for all $a,a_1,a_2\in A$ and all $\lambda\in[0,1]$.
\end{defn}

\section{Characterisation of encodings}\label{sect characterise}

Complex numbers and non-Hermitian operators have become integral to the study of quantum mechanics. However, it is much harder to make well-motivated assumptions about simulations in these cases.
Therefore, to avoid over asserting, the encoding, $\gamma: A \mapsto B$, acts only between JB algebras where elements correspond to observables and therefore properties of maps on these objects can be motivated physically.
However, a finite dimensional special JB algebra $A$ ( resp. $B$) is necessarily contained in an associative algebra $\mathcal{A}$ ( resp. $\mathcal{B}$).
Our first step is to connect to more familiar associative algebras (such as $C^{*}$-algebras) by extending the encoding to the encompassing associative algebras, $\Gamma : \mathcal{A}\mapsto \mathcal{B}$.
The only assumption on this extended map is that its action on the self-adjoint elements of the associative algebra (the observables) is that of an encoding as described in \cref{defn sim map}.
A priori the action of the extended map on non-symmetric elements under the involution is unconstrained.



In what follows, $A$ and $B$ are finite-dimensional special JB algebras.
We have motivated algebras of subsets observables and separately that simulations should be spectrum preserving, our first result demonstrates these two ideas are strongly connected mathematically.
\cref{babylemma3} demonstrates this connection between spectrum-preserving maps and Jordan homomorphisms that is key to then leverage results from the literature.

We then show that extended encodings are uniquely determined by the original encoding by combining results from the associative algebra literature, \cite{jordanoalgebrasHancheStormer, mccrimmon, jordanVNWigner, Martindale, JacobsonandRickart}.
Therefore, considering the mathematical form of $\Gamma$ instead of $\gamma$ introduces no ambiguity (\cref{lm:extended}).
The first case considered is that of a encoding $\gamma: A \mapsto B$ where $B$ is a semisimple JB algebra.
This includes the case where all observables are simulated when $A \cong \text{Herm}_n(\mathbb{F})$.

\begin{thm}[Characterisation of encodings of simple algebras]\label{thm: simple map charac}
    Given an encoding $\gamma: A \rightarrow B$, where $A$ is a simple JB algebra, and $B$ is a semisimple algebra, the extended encoding $\Gamma: \mathcal{A}\rightarrow \mathcal{B} $ is  (up to unitary equivalence) of the form
    \[ \Gamma(\alpha)= \oplus_{i} \Gamma^{p_{i},q_{i}}(\alpha)= \oplus_{i} (\alpha^{\oplus p_{i}} \oplus \bar{\alpha}^{\oplus q_{i}}) \]
    for all $\alpha\in\mathcal{A}$ and $p_{i},q_{i} \in \mathbb{N}_{0}$ such that $p_{i}+q_{i} = m_{i}/n$ where $\textup{dim}(\mathcal{A}) = n$, $\textup{dim}(\mathcal{B}_{i}) = m_{i}$.
\end{thm}

The resulting form reduces to that found in \cite{Cubitt2019} when $B$ is a simple algebra.
When $B$ is simple, the semisimple algebra $A$ can be embedded into different ideals of $B$. 
Recall that semi-simple algebras arise in quantum mechanics when there are superselection rules in the system \cite{supersel}, that is, rules which do not allow certain superpositions.
In such systems, different superselection sectors correspond to different summands in the algebra.
An important application of superselection rules is in systems with indistinguishable particles.
Here, the algebra of observables is a semisimple algebra, which can sometimes be decomposed into a direct sum of the algebra of fermions and the algebra of bosons (for an overview, see \cite{fermionsforbabies}).

\begin{col}[Characterisation of encodings of semisimple algebras]\label{col: semisimple}
     Given an encoding $\gamma: A \rightarrow B$, the extended encoding $\Gamma: \mathcal{A}\mapsto \mathcal{B}$ is completely characterised (up to unitary equivalence) by
    \[\Gamma(\alpha)= \left(\bigoplus_{j} ( \alpha_{j})^{\oplus p_{ij}} \oplus ( \bar{\alpha}_{j})^{\oplus q_{ij}} \right) \]
    for all $\alpha= \alpha_{1} + ...+ \alpha_{i}+ ...+\alpha_{m}\in\mathcal{A}$ in $\mathcal{A} \cong \mathcal{A}_{1} \oplus ...\oplus \mathcal{A}_{i} \oplus ... \oplus \mathcal{A}_{m}$ and $\mathcal{B} \cong \mathcal{B}_{1} \oplus ... \oplus \mathcal{B}_{i} \oplus ... \oplus \mathcal{B}_{n}$.
    For some $\{p_{ij}\},\{q_{ij}\} \in \mathbb{N}_{0}$ such that $ (p_{ij}+q_{ij}) n_i= m_j$ where $\textup{dim}(\mathcal{A}_{i}) = n_{i}$, $\textup{dim}(\mathcal{B}_{i}) = m_{i}$, and the integers $p_{ij},q_{ij}$ denote the multiplicities of the embeddings of the $j$th summand of $\mathcal{A}$ into the $i$th summand of $\mathcal{B}$.
\end{col}

The key difference between \cref{thm: simple map charac} and \cref{col: semisimple} is that when both algebras are semisimple,  each summand of the system of interest may be embedded in different summands of  $B$ with different multiplicities).
Hence every superselection sector of the original algebra may be sent to a different superselection sector in $\mathcal{B}$.
For example, if one were to simulate indistinguishable particles (which are semisimple due to superselection rules) then the simulation may map different degeneracies of the fermionic and bosonic sectors.

In most constructive simulations the freedom to take multiple copies of both the domain system and its complex conjugate is not used.
To our knowledge, the only example of a constructive simulation that utilises the freedom to take such copies is the simulation of complex Hamiltonians with real Hamiltonians (\cite[Lemma 7]{Cubitt2019}).
Therefore, the additional freedom discovered here of taking different degeneracies for different subalgebras seems to have only artificial benefit.
This does however, highlight underexplored possibilities of constructive simulations and we hope that outlining the possible mathematical parameters of simulation maps aids the search for new constructions.

\subsection*{Example: Fermionic encodings}

Simulating systems of fermions is of crucial importance for fields such as condensed matter and quantum chemistry, where electronic wavefunctions govern material properties and molecular interactions \cite{Stanisic2022, BravyiKitaev}.
Fermionic systems are often\footnote{They can be equivalently defined by self-adjoint operators $p_{i},q_{j}$ in a unital associative algebra with involution satisfying $p_{k}p_{j}+p_{j}p_{k}= 2 \delta_{jk} \mathds{1}$ and $p_{k}q_{j}+q_{j}p_{k}=0$ such that $a_{k}= \frac{1}{2}(q_{k}+ip_{k})$.} defined through annihilation and creation operators $a_{k}$ and $a^{*}_{k}$ obeying anticommutation relations: 
\begin{equation}
a_{j}^{*}a_{k} + a_{k}a_{j}^{*} = 2 \delta_{jk}\mathds{1} \quad \text{and} \quad a_{j}a_{k}+a_{k}a_{j}= 0,
\end{equation}
The algebra of fermionic observables is then defined by the universal $C^{*}$-algebra generated by the fermionic creation and annihilation operators satisfying the canonical anticommutation relations -- the CAR algebra.
For both analogue and digital quantum simulation, an encoding is required to translate the fermionic Hamiltonian from these fermionic operators to qubit operators.
We will take this subalgebra of observables to consider the implications of the above characterisations\footnote{While in the previous section we extended the notion of encodings from special JB algebras to the $C^{*}$-algebras that they are contained in.
Conversely, it is straightforward to verify that, given a finite-dimensional $C^*$-algebra, the set of self-adjoint elements is always a special JB algebra.}


Several encodings of the CAR algebra into qubit systems have been constructed \cite{JordanWigner,BravyiKitaev, fermiontoqubitcompact, VerstraeteCirac2005, }. 
As the CAR algebra is known to be simple\footnote{The CAR algebra contains a set of matrix units, implying that the algebra is a simple $C^{*}$-algebra isomorphic to the simple algebra $M_{2^{n}}(\mathbb{C})$ via a canonical embedding into a matrix ring.} \cite{arvesoninvitation,JacobsonandRickart,rowenringtheory} all encodings acting on full CAR must be of the form of \cref{thm: simple map charac}.
In fact all the aforementioned encodings take a particularly simple case of \cref{thm: simple map charac} whereby $p=1$ and $q=0$, so the nuances are in the form of the unitary transformation -- hence all proposed fermionic encodings are unitarily equivalent.

The number of particles in a fermionic state is given by the fermion-number and physical states obey parity superselection rules: superpositions of even fermion number states and odd fermion number states are forbidden. 
This can be understood by considering locality \cite{FermionsforQuantumInfoPeople}.
By extension, physical fermionic operators preserve the fermion-number parity.
Therefore, the Hamiltonians describing the physical systems we wish to simulate on a quantum computer consist of terms with an even number of creation/annihilation operators.

Fermionic operators preserving fermion-number parity are described by the \emph{even subalgebra of the CAR algebra}, dubbed the `physically relevant algebra' \cite{bravyikitaevforelectrons,BravyiKitaev}\footnote{Since the CAR algebra is simple, the parity operator clearly does not diagonalise the algebra and indeed the operators describing the odd fermionic operators that alter the fermion-number parity do not describe a closed subalgebra.}.
Since the even CAR subalgebra is itself a $C^{*}$-algebra, we can also consider directly encoding this subalgebra.
In infinite dimensions, the even subalgebra is isomorphic to the full algebra of fermions \cite{stormer1970even}.
However, in the finite-dimensions, this is not the case and the encodings of the physically relevant even fermionic subalgebra are characterised by \cref{col: semisimple}.

Consider the automorphism $\gamma$ on the CAR algebra which acts as $\gamma: a_{i} \mapsto -a_{i}$ for every generator and extend multiplicatively by defining
\begin{equation}
\gamma(a_{i_1}a_{i_2}\cdots a_{i_k}) = (-1)^k\, a_{i_1}a_{i_2}\cdots a_{i_k}.
\end{equation}
The even algebra is invariant under the action of $\gamma$.
In a concrete representation, the automorphism $\gamma$ is implemented by a self-adjoint unitary parity operator $P$, satisfying
\begin{equation}
 \gamma(a) = P\, a\, P  \quad \text{for all } a.
\end{equation}
Since $P$ is self-adjoint and unitary, we may define the central projections
\begin{equation}
E_{+} = \frac{1+P}{2} \quad \text{and} \quad E_{-} = \frac{1-P}{2},
\end{equation}
Thus, the even algebra $\mathcal{A}_{even}$ splits as an algebra according to
\begin{equation}
\mathcal{A}_{even} = E_{+}\,\mathcal{A}_{even} \oplus E_{-}\,\mathcal{A}_{even}.
\end{equation}
 where both $E_{+}\,\mathcal{A}_{even}$ and $E_{-}\,\mathcal{A}_{even}$ are subalgebras of $\mathcal{A}_{even}$\footnote{That a complete set of orthogonal central idempotents partitions an algebra into its simple components is a standard argument in noncommutative algebra, see for example \cite[Chapter 1]{noncommutativealgebra} or \cite[Part 1 Chapter 2]{garling2011}.}.
 Using well-known theorems in the representation of finite-dimensional algebras (see \cite[Theorem 6.6.1]{garling2011}), it follows that
\begin{equation}
A_{even} \cong M_{2^{n-1}}(\mathbb{C}) \oplus M_{2^{n-1}}(\mathbb{C}).
\end{equation}
Therefore, the even alegbra is semisimple and encodings of just the parity preserving fermionic observables have the additional freedoms characterised by \cref{col: semisimple}.

\subsection*{Acknowledgements}

We thank Robert Evans for helpful comments on an earlier version of the manuscript. MA thanks Niklas Galke for helfpul discussions. HA is supported by EPSRC DTP Grant Reference:
EP/N509577/1 and EP/T517793/1. HA and TSC were supported in part by the EPSRC Prosperity Partnership in Quantum Software for Simulation and Modelling (grant EP/S005021/1), and by
the UK Hub in Quantum Computing and Simulation, part of the UK National Quantum
Technologies Programme with funding from UKRI EPSRC (grant EP/T001062/1).  MA is supported by the Mexican National Council of Science and Technology
(CONACYT), by EPSRC, and by
the Simons Foundation It from Qubit Network.

\subsection*{Proofs}

\begin{figure}[h!]
\centering
\begin{tikzpicture}
\begin{scope}[on background layer]
         \node [inner sep=0pt] (1) at (0, 0) {\includegraphics[trim={0cm 0cm 0cm 0cm},clip,width=0.85\linewidth]{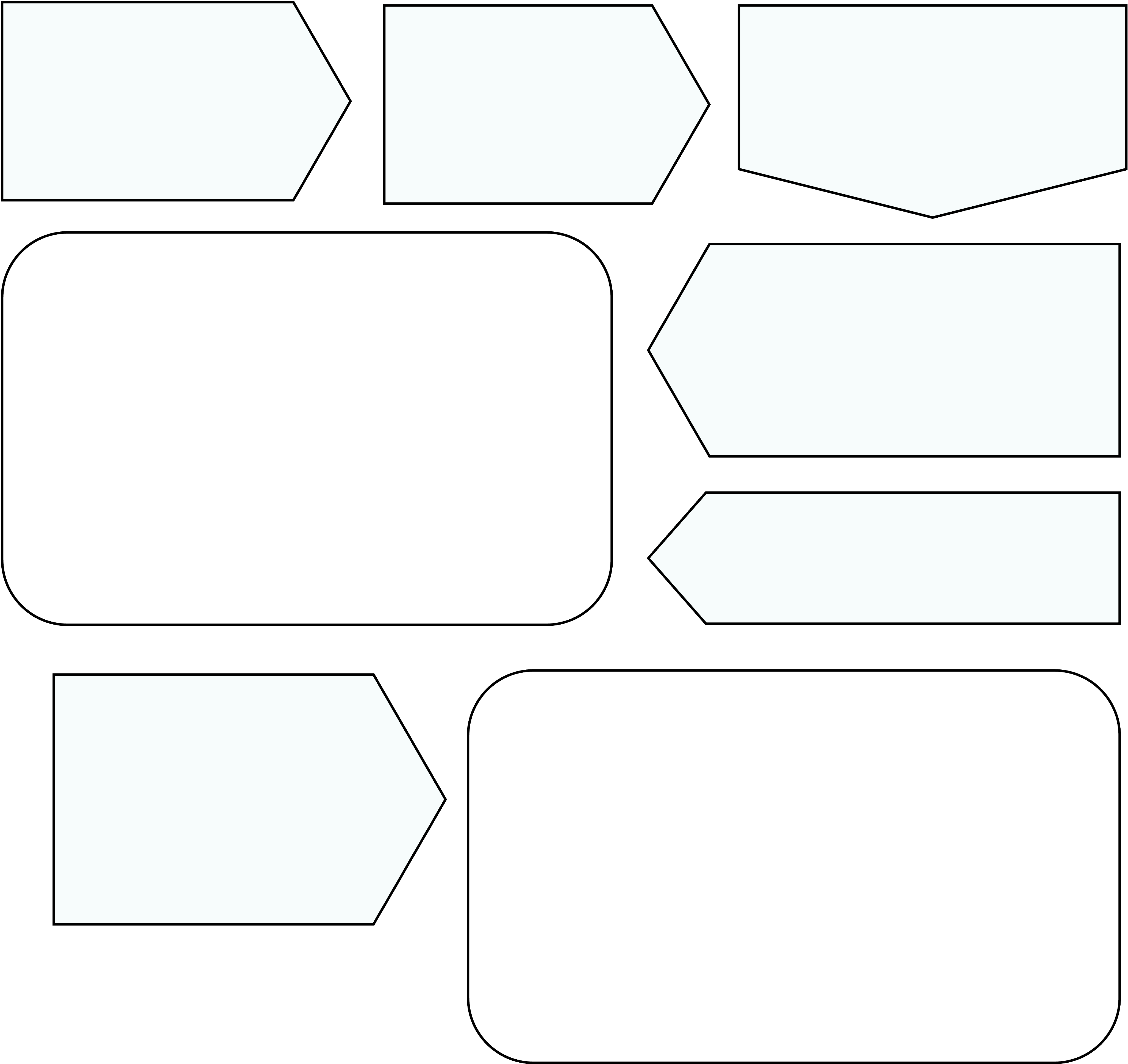}};
\end{scope}
		\node [align=left, anchor=west] (0) at (-6.4, 5.6) {\scriptsize Simulation map between };
		\node [align=left, anchor=west] (0) at (-6.4, 5.2) {\scriptsize subsets of observables, };
		\node [align=left, anchor=west] (0) at (-6.4, 4.8) {\scriptsize $\gamma: A \mapsto B$, where $A$};
		\node [align=left, anchor=west] (0) at (-6.4, 4.4) {\scriptsize is simple.};
		\node [align=left, anchor=west] (0) at (-2.1, 5.2) {\scriptsize $\gamma$ is shown to be a Jordan };
		\node [align=left, anchor=west] (0) at (-2.1, 4.8) {\scriptsize homomorphism using};
  \node [align=left, anchor=west] (0) at (-2.1, 4.4) {\scriptsize \cref{babylemma3}.};
		\node [align=left, anchor=west] (0) at (2, 5.6) {\scriptsize  \cref{lm:extended} extended simulation };
		\node [align=left, anchor=west] (0) at (2, 5.2) {\scriptsize map  $\Gamma: \mathcal{A}\mapsto \mathcal{B}
	$ shown to be an};
	\node [align=left, anchor=west] (0) at (2, 4.8) {\scriptsize associative homomorphism and };
	\node [align=left, anchor=west] (0) at (2, 4.4) {\scriptsize uniquely determined by $\gamma$.};
	\node [align=left, anchor=west] (0) at (1.8, 2.8) {\scriptsize $\Gamma(\alpha) = U' (\Gamma_+(\alpha)\oplus \Gamma_-(\alpha))U'^*$ };
	\node [align=left, anchor=west] (0) at (1.5, 2.3) {\scriptsize for all $\alpha \in \mathcal{A}$ where $\Gamma_{+(-)}$ is (anti)-};
	\node [align=left, anchor=west] (0) at (1.5, 1.9) {\scriptsize -linear over $\mathbb{C}$ and $U\in\mathcal{B}$ where};
	\node [align=left, anchor=west] (0) at (1.5, 1.5) {\scriptsize $U U^* = \mathds{1}$.};
	\node [align=left, anchor=west] (0) at (1.5, 0) {\scriptsize \cite{cstaralgebrasbook} characterisation of };
	\node [align=left, anchor=west] (0) at (1.5, -0.4) {\scriptsize $*$-homomorphisms in finite dim. };
	\node [align=left, anchor=west] (0) at (-5.8, -2.2) {\scriptsize Then if $A$ semisimple the };
	\node [align=left, anchor=west] (0) at (-5.8, -2.6) {\scriptsize extended homomorphism a };
	\node [align=left, anchor=west] (0) at (-5.8, -3) {\scriptsize direct sum of of the simple case  };
	\node [align=left, anchor=west] (0) at (-5.8, -3.5) {\scriptsize$\mathrm{Hom}(\bigoplus M_i, \mathcal{B}) \cong \bigoplus_i (M_i, \mathcal{B}) $  };
	\node [align=left, anchor=west] (0) at (-5.8, -4) {\scriptsize so $\Gamma(\alpha) = U' \bigoplus_i \Gamma_i (\alpha _i )U'^*$  };
	\node [align=left, anchor=west] (0) at (-6.3, 3) {\scriptsize \cref{thm: simple map charac}: if $A$ is simple, the extended};
	\node [align=left, anchor=west] (0) at (-6.3, 2.6) {\scriptsize simulation map is of the form };
	\node [align=left, anchor=west] (0) at (-6.1, 1.1) {\scriptsize $\Gamma^{p,q}(\alpha)= U\begin{pmatrix}\alpha &  & &&&\\ & \ddots  & &&&\\  & & \alpha&&&\\ &&&\bar{\alpha}&&\\ &&&&\ddots&\\ &&&&&\bar{\alpha}\end{pmatrix}U^{*}$ };
	\node [align=left, anchor=west] (0) at (-6.1, -0.5) {\scriptsize with $p$($q$) copies of $\alpha$$(\bar{\alpha})$ and $U U^* = U^* U =\mathds{1}_{\mathcal{B}}$. };
	\node [align=left, anchor=west] (0) at (-1, -2.1) {\scriptsize \cref{col: semisimple}: if $A$ is semisimple, the extended };
	\node [align=left, anchor=west] (0) at (-1, -2.5) {\scriptsize simulation map is of the form };
	\node [align=left, anchor=west] (0) at (-1, -4) {\scriptsize $\Gamma^{\{p_i\},\{q_i\}}(\alpha)= U\begin{pmatrix}\alpha_1 &  & &&&\\ & \ddots  & &&&\\  & &\alpha_i&&&\\ &&&\bar{\alpha_1}&&\\ &&&&\ddots&\\ &&&&&\bar{\alpha_i}\end{pmatrix}U^{*}$ };
	\node [align=left, anchor=west] (0) at (-1, -5.6) {\scriptsize with $p_i$($q_i$) copies of $\alpha_i$$(\bar{\alpha_i})$ where $U U^* = U^* U =\mathds{1}_{\mathcal{B}}$. };
\end{tikzpicture}
\caption{Outline of proof idea for \cref{thm: simple map charac} and \cref{col: semisimple}.}
 \label{fg extended proof idea}
\end{figure}

This section contains the proofs of \cref{thm: simple map charac} and \cref{col: semisimple}.
\cref{fg extended proof idea} gives an overview of the roadmap to these results. 

The first step is to show that the extended simulation map acting between associative algebras is uniquely determined by the simulation map acting between simple JB algebras.
First note that any element of the Jordan algebra has a spectral decomposition since,

\begin{lemma}\label{spectral} [\cite[Lemma 2.9.4]{jordanoalgebrasHancheStormer}] Let $A$ be a finite-dimensional, formally-real, unital Jordan algebra over $\mathbb{R}$. Then any element of $A$ is contained in some associative subalgebra of $A$, and every such subalgebra is of the form $\mathbb{R}p_{1} \oplus \mathbb{R}p_{2}...\oplus \mathbb{R}p_{n}$, where the $p_{i}$ are a set of pairwise orthogonal minimal idempotents with sum $1$. 
\end{lemma}

We use this to first show a connection between spectrum preserving maps between Jordan algebras and Jordan homomorphisms.

\begin{lemma}\label{babylemma3} Let $A$ and $B$ be two finite-dimensional special JB algebras. Let $\gamma: A \rightarrow B$ be a spectrum-preserving, $\mathbb{R}$-linear map. Then $\gamma$ is a Jordan homomorphism. 
\end{lemma}
\begin{proof}
     By \cref{spectral}, every element $a \in A$  can be written as $ \sum_{i} \lambda_{i} p_{i}$ with $\lambda_{i} \in \mathbb{R}$ and $p_{i}$ orthogonal projections. Observe now that $p_{i}$ is a projection if and only if $ \textup{spec}(p_{i}) \in \{ 0,1\}$. Let $p_{i}$ and $p_{j}$ be a pair of orthogonal projections in $O$. Then $p_{i} + p_{j}$ is also a projection. Since $\gamma$ is spectrum preserving, it follows that $\gamma(p_{i}+p_{j})$ is also a projection, and therefore $\gamma(p_{i}+p_{j})= (\gamma(p_{i}+p_{j}))^{2}$. Using linearity on the right hand side gives $\gamma(p_{i})\gamma(p_{j}) + \gamma(p_{j})\gamma(p_{i}) =0$. And hence for every $a \in A$ $\gamma(a^{2})= \gamma(a)^{2}$, so   $\gamma$ is a Jordan homomorphism.  
\end{proof}

In this section we will be discussing simple and semisimple Jordan algebras therefore recall for completeness the classification theorem of formally real unital Jordan algebras.

\begin{thm}\label{JordanvnWigner} Every finite-dimensional, formally real, unital Jordan algebra $A$ is a direct sum of simple algebras. If $A$ is simple, then it contains $n \geq 1$ pairwise orthogonal and strongly connected minimal idempotents with sum $1$.  If $n=1$, then $A \cong \mathbb{R}$. If $n=2$, then $A$ is a spin factor. If $n \geq 3$, $A$, then $A$ is special and it is isomorphic to one of $H_{n}(\mathbb{R})$, $H_{n}(\mathbb{C})$, or $H_{n}(\mathbb{H})$. If $n=3$, then $A$ is isomorphic to the exceptional Jordan algebra $H_{3}(\mathbb{O})$.  
\begin{proof}
    For a proof, see \cite[2.9.6]{jordanoalgebrasHancheStormer}, and \cite{mccrimmon}. For a proof that $H_{3}(\mathbb{O})$ is exceptional, see \cite[Corollary 2.8.5]{jordanoalgebrasHancheStormer} and \cite{jordanVNWigner}.  
\end{proof}
\end{thm}

We can now consider the extended simulation map between the associative algebras, which can be chosen to be C$^*$-algebras (see \cref{appen C*}).
We will first show that these extended simulation maps are uniquely determined by the maps between Jordan algebras using \cref{thm our martindale}.

\begin{thm}\label{thm our martindale}[\cite{Martindale} theorem 2 and \cite{JacobsonandRickart} theorem 4]
    Let $\mathcal{A}$ be a ring with canonical involution $*$ which contains $n$ orthogonal symmetric idempotents $e_{1}, ....e_{n}$. Let $A= \{ a \in \mathcal{A} : a^{*}=a \}$. Then any Jordan homomorphism of $A$ into an arbitrary ring $\mathcal{B}$ can be extended uniquely to an associative homomorphism of $\mathcal{A}$ into $\mathcal{B}$. 
\end{thm}

The theorem above allows us to prove the following result. 

\begin{lemma}[Extended simulation for simple algebras]\label{lm:extended}
    Consider a simulation map $\gamma: A \rightarrow B$ where $A$ are is a simple Jordan algebra and let $\mathcal{A}$ ( resp. $\mathcal{B}$) be the associative algebra in which $A$ ( resp. $B$) is contained.
    The extended simulation map $\Gamma: \mathcal{A}\rightarrow \mathcal{B}$ is uniquely determined by $\gamma$.
    Additionally for all $\alpha_1, \alpha_2 \in \mathcal{A}:$
    \begin{enumerate}[(i.)]
         \item $\Gamma(\alpha_1+\alpha_2)= \Gamma(\alpha_1)+ \Gamma(\alpha_2)$;\label{item add}
        \item $\Gamma(\alpha_1 \cdot \alpha_2)= \Gamma(\alpha_1) \Gamma(\alpha_2)$;\label{item multiply}
        \item $\Gamma(\mathds{1}_{\mathcal{A}})= \mathds{1}_{\mathcal{B}}$;\label{item unital}
        \item $\Gamma(\lambda\alpha_1) = \lambda \Gamma(\alpha_1)$ for $\lambda \in \real$;\label{item reals}
        \item $\Gamma(\alpha_{1}^{*})= \Gamma(\alpha_{1})^{*}$.\label{item inv}
    \end{enumerate}
\end{lemma}
\begin{proof}
Spectrum preserving implies $\gamma(0) = 0$. We will now show that this, together with convexity, implies $\mathbb{R}$-linearity. 
For any $\lambda <0$ define $p: = \frac{\lambda}{\lambda - 1} \in [0,1]$ and for an element $a\in A$ let $c = \frac{p}{(p-1)}a = \lambda a \in A$.
Then 
\begin{align}
  \gamma (p a + (1-p) c) & = \gamma (0) = 0\\
  & = p \gamma(a) + (1-p)\gamma(\lambda a).
\end{align}
Therefore $\lambda \gamma(a) = \gamma(\lambda a)$.
Repeating this logic starting with $\{a' =\lambda a, c' =\lambda^2 a\}$ gives $\lambda^2 \gamma(a) = \gamma (\lambda^{2} a)$ and homogeneity for all $\lambda \in \mathbb{R}$. 
Combining convexity with homogeneity gives $\mathbb{R}$-linearity.
Hence by \cref{babylemma3} $\gamma$ is a Jordan homomorphism. 
By \cref{JordanvnWigner}, $A$ contains $n$ projections, and therefore by \cref{thm our martindale}, there exists a unique extension to an associative homomorphism $\gamma: \mathcal{A}\mapsto \mathcal{B}$, which satisfies \cref{item add}, and \cref{item multiply} automatically.
Unitality (\cref{item unital}) also follows from \cite{Martindale} and \cite{JacobsonandRickart}\footnote{It is worth noting that every spectrum preserving map $\gamma$ between Banach algebras can be chosen to be unital. This follows from the fact that every spectrum preserving map $\varphi$ is invertibility preserving:  so that the element $\varphi(1)$ always has an inverse, which allows us to define the unital map $a \mapsto \varphi^{-1}(1)\varphi(a)$ for any map $\varphi$. }.

\cref{item reals} follows from real linearity of $\gamma$ and \cref{item multiply}: for all $\alpha\in \mathcal{A}$, $\Gamma(\lambda\mathds{1} \alpha) = \gamma(\lambda\mathds{1})\Gamma(\alpha) = \lambda \Gamma(\alpha)$.
It remains to show \cref{item inv}.
One can verify that $\gamma(a^{*})=\gamma(a)^{*}$ directly from linearity and the fact that $\gamma$ preserves projections, and since $\forall a \in A$ $\gamma(a)=\Gamma(a)$, the statement is true for self-adjoint elements. 
It remains to verify the action of $\Gamma$ on general elements of $\mathcal{A}$, but since every element is of the form $a+ib$ for some self adjoint elements $a,b$, it is enough to check that $\Gamma((i\mathds{1})^{*})=\Gamma(i\mathds{1})^{*}$. We will do this by showing that $\Gamma(\alpha^{*})=\Gamma(\alpha)^{*}$ for any element which satisfies $\alpha^{*}=\alpha^{-1}$. 
Recall first that since $\Gamma$ is unital we necessarily have $\Gamma(\alpha^{-1})=\Gamma(\alpha)^{-1}$, for all invertible elements. Hence, for any unitary, we have $\Gamma(\alpha^{-1})=\Gamma(\alpha^{*})=\Gamma(\alpha)^{-1}$. We also have $\Gamma(\alpha)=\Gamma(\alpha^{*})^{-1} \implies \Gamma(\alpha)^{*}=\Gamma(\alpha^{*})^{*-1}$. Similarly, we have $\Gamma(\alpha^{*})^{*}=\Gamma(\alpha)^{-1*} \implies \Gamma(a)^{-1}=\Gamma(a^{*})^{*-1}$. 
Hence $\Gamma((i\mathds{1})^{*})=\Gamma(i\mathds{1})^{*}$. 

\end{proof}

In order to show that any extended simulation map for associative algebras is characterised by a Jordan homomorphism, we will utilise the following well-known result from the theory of $*$-algebras.

\begin{lemma}[Characterisation of finite dimensional $*$-homomorphisms]\label{lm copies of id}[See \cite{cstaralgebrasbook} Corollary III.1.2.]
If $\pi$ is a unital $*$-homomorphism of a finite dimensional C$^*$ algebra, $\mathcal{A}$, there is an integer $p$ such that $\pi$ is unitarily equivalent to $p$ copies of the identity representation of $\mathcal{A}$.
\end{lemma}

Using \cref{lm:extended} and \cref{lm copies of id} we can now characterise the extended simulation map when $A$ and $\mathcal{A}$ are simple. 

\begin{repthm}{thm: simple map charac}[Characterisation of simple simulation map]
    Given a simulation map $\gamma: A \rightarrow B$, where $A$ is a simple algebra, and $B$ is a semisimple algebra, the extended simulation map $\Gamma: \mathcal{A}\rightarrow \mathcal{B}$ is (up to unitary equivalence) of the form
    \[  \Gamma (\alpha)= \oplus_{i} \Gamma^{p_{i},q_{i}}(\alpha)= \oplus_{i} (\alpha^{\oplus p_{i}} \oplus \bar{\alpha}^{\oplus q_{i}}) \] 
    for all $\alpha\in\mathcal{A}$ and $p_{i},q_{i} \in \mathbb{N}_{0}$ such that $p_{i}+q_{i} = m_{i}/n$ where $\textup{dim}(\mathcal{A}) = n$, $\textup{dim}(\mathcal{B}_{i}) = m_{i}$. 
\end{repthm}
\begin{proof}
    We now show that $\Gamma$ is an extended simulation map $\iff$ $\Gamma(\alpha)= U(\alpha^{\oplus p} \oplus \bar{\alpha}^{\oplus q}) U^{\dagger}$ for some $p,q \in \mathbb{N}_{0}$. 
    The $\impliedby$ direction can be verified directly. 

    We now consider the $\implies$ direction.
    The action of $\Gamma$ on the self-adjoint elements is determined by $\gamma$, so it remains to define the action on general elements of $\mathcal{A}$.
    Since every element of $\mathcal{A}$ can be written as $ a + ib$ for some $a,b \in A$, it is enough to infer the action of $\Gamma$ on $i \mathds{1}$.
    Observe that $\Gamma(i \mathds{1})^{2}= -\mathds{1}$, so the operator $\Gamma(i \mathds{1})$ has spectrum $\{\pm i\}$.
    Since $\Gamma(i \mathds{1})$ is normal and central, it can be written as a sum of central projections (see \cite{lncstaralgebras} lemma 1.7.4), and hence $\Gamma(\alpha)=  \Gamma_{+}(\alpha) \oplus \Gamma_{-}(\alpha)$ where the $\pm$ signs denote the action on the two invariant eigenspaces. 
    
    \cref{lm:extended} gives that $\Gamma_{+}$ is a linear map over the complex numbers and therefore a $*$-homomorphism, hence \cref{lm copies of id} applies directly to $\Gamma_+$, so $\Gamma_+(\alpha) = U_+^\dagger \alpha^{\oplus p} U_+$ for some unitary $U_+$ and integer $p$. We have also shown that $\Gamma_{-}$ is an anti-linear map. Using that any anti-linear map can be written as the complex conjugate of a linear map, we write $\Gamma_{-}=\bar{\varphi}$ for some linear $\varphi$. Since $\varphi$ is linear, by \cref{lm copies of id}: $\varphi = U_-^\dagger \alpha^{\oplus q} U_- $ and hence $\Gamma_{-}(\alpha) = U_-^\dagger \bar{\alpha}^{\oplus q} U_- $ for some unitary $U_-$ and integer $q$.
    In the case where 
    $A$ and $B$ are simple algebras, if $ \mathcal{B}=\mathrm{Im}(\Gamma)$ and $\mathcal{B}$ is simple, it is clear that $\Gamma(\alpha)= U^\dagger\alpha U$ or $\Gamma(\alpha) = U^\dagger\bar{\alpha} U$ for all $\alpha \in \mathcal{A}$  \footnote{It is worth noting that this characterisation of spectrum-preserving and linear maps which are also surjective can be given in infinite dimensions (see \cite{spectrumpreserving})}. Else 
    $\Gamma(\alpha)= U \alpha^{\oplus^{p}} \oplus \bar{\alpha}^{\oplus q} U^{\dagger}$ by \cref{lm copies of id}. 
    For the case where $\mathcal{B}$ is a semi-simple algebra, note that $\mathrm{Hom}(N,\oplus M_{i}) \cong \oplus_{i} \mathrm{Hom}(N,M_{i})$. If $\mathcal{B}= \oplus \mathcal{B}_{i} $, denote by $\epsilon_{i}: \mathcal{B} \rightarrow \mathcal{B}_{i}$ the canonical surjection onto the $i$th summand of $\mathcal{B}$. Then the map $ \Gamma_{i} = \epsilon_{i} \circ \Gamma : \mathcal{A} \rightarrow \mathcal{B_{i}} $ is a unital map between simple algebras, and $\Gamma = \Gamma_{i} \oplus ...\oplus \Gamma_{i}$, with $ \Gamma_{i}= U \alpha^{p_{i}} \oplus \bar{\alpha}^{q_{i}} U^{\dagger}$. 
\end{proof}

As a corollary, we obtain a characterisation in the case where $A$ is semisimple.

\begin{repcol}{col: semisimple}[Characterisation of simulation map between semi-simple algebras]
    Given a simulation map $\gamma: A \rightarrow B$, any extended simulation map $\Gamma: \mathcal{A}\mapsto \mathcal{B}$ is  (up to unitary equivalence) of the form
    \[\Gamma(\alpha)= \left(\bigoplus_{j} ( \alpha_{j})^{\oplus p_{ij}} \oplus ( \bar{\alpha}_{j})^{\oplus q_{ij}} \right) \]
    for all $\alpha= \alpha_{1} + ...+ \alpha_{i}+ ...+\alpha_{m}\in\mathcal{A}$ in $\mathcal{A} \cong \mathcal{A}_{1} \oplus ...\oplus \mathcal{A}_{i}... \oplus \mathcal{A}_{m}$ and $\mathcal{B} \cong \mathcal{B}_{1} \oplus ... \oplus \mathcal{B}_{i} ... \oplus \mathcal{B}_{n}$. 
    For some $\{p_{ij}\},\{q_{ij}\} \in \mathbb{N}_{0}$ such that $ (p_{ij}+q_{ij}) n_i= m_j$ where $\textup{dim}(\mathcal{A}_{i}) = n_{i}$, $\textup{dim}(\mathcal{B}_{i}) = m_{i}$, and the integers $p_{ij},q_{ij}$ denote the multiplicities of the embeddings of the $j$th summand of $\mathcal{A}$ into the $i$th summand of $\mathcal{B}$. 
\end{repcol}
\begin{proof}
By the same argument as \cref{lm:extended}, $\gamma$ is a Jordan homomorphism. We now use the vector space structure of the Jordan algebras $A,B$ as well as the fact that for arbitrary $R$-modules $M,N$, there is an isomorphism $\textup{Hom}(\oplus A_{i},N) \cong  \oplus_{i} \textup{Hom}(A_{i},N)$ (see \cite{noncommutativealgebra}). We can therefore write $\gamma = \oplus_{i} \gamma_{i} $ where $\gamma_{i}= \gamma \vert_{A_i}(a) : A_i \mapsto B$ denotes the restriction of $\gamma$ to the ith summand of $A$. Since each $\gamma_{i}$ is a Jordan homomorphism from a simple algebra, we may apply \cref{thm: simple map charac}. 
It follows that there exists an extended simulation map $\Gamma_{i}$ for each restriction $\gamma_{i}$ and combining with \cref{thm: simple map charac} to extend to the case where both $A$ and $B$ are semisimple gives the desired result. 
\end{proof}

\printbibliography

\begin{appendices}
    \section{C$^*$-algebras and Jordan algebras}\label{appen C*}

    In the main text, we state that the subspace of self-adjoint elements of an associative algebra $\mathcal{A}$ with involution is a special Jordan algebra, and that given a set of observables, one can generate a special Jordan algebra by taking linear combinations and Jordan products in the set. Here we justify this claim as well as outline how the Jordan algebras discussed in the text can be embedded into $C^{*}$ algebras. 
    
    \begin{proposition}\label{SPJAus} Let $\mathcal{A}$ be an associative algebra with involution. The set $A$ of $*$-symmetric elements of $\mathcal{A}$ (in the sense that $a^{*}=a$) is a special Jordan algebra. 
    \end{proposition}
\begin{proof}
 We must show that if $a,b \in A$, then $a \circ b \in A$. If $a,b \in \mathcal{S}(\mathcal{U}, J)$, then $a \circ b = ab+ba = (a \circ b)^{*}$, so the set is indeed closed under the Jordan product. 
\end{proof}

    \begin{col}\label{MotherAlgebra} Let $\mathcal{a}$ be an associative algebra over $\mathbb{R}$, and suppose that $\mathcal{A}$ possesses a linear involution $*$ such that $\forall a \in \mathcal{A}$ and $\forall \lambda \in \mathbb{R}$ $(\lambda a)^{J}= \lambda a^{J}$. Let $A$ denote the set of $*$-symmetric elements of $\mathcal{A}$ in the sense that $a^{*}=a$. Let $X=\{  a_{1}, ... a_{n} \}$ for some finite $n$, and where $a_{i} \in A$. Let $J<a_{i}>$ denote the set of all Jordan products and real linear combinations of the elements in $X$. Then $J<a_{i}>$ is a special Jordan algebra. 
\end{col}
\begin{proof}
    Follows directly from \ref{SPJAus}. 
\end{proof}

We further claim to connect the special formally real algebras to the more familiar $C^{*}$ algebras.
This can be jusitified by first recalling the the well-known classification theorem of finite-dimensional formally real Jordan algebras \cref{JordanvnWigner}.

Note that the algebras considered in this paper exclude the case $n=3$. For $n=1$, it is clear that $\mathbb{R}$ can be embedded in the $C^{*}$ algebra $\mathbb{C}$. For $n \geq 3$, and $A \cong H_{n}(\mathbb{R})$ or $A \cong H_{n}(\mathbb{C})$, the embedding is also clear. For $n \geq 3$, and $A \cong H_{n}(\mathbb{H})$, one can identify $\mathbb{H}$ with a real $*$ subalgebra of $M_{n}(\mathbb{C})$ via the representation 
\begin{equation}
    a + bi + cj + dk \mapsto \begin{pmatrix}
        a + bi & c-di \\ 
        -c-di & a-bi
    \end{pmatrix}
\end{equation}
For $n=2$, the isometric embedding into a $C^{*}$ algebra is more complicated, so we refer the reader to \cite{jordanoalgebrasHancheStormer}. 

Although we have so far focused on embedding formally real algebras into associative $C^{*}$-algebras, it is also straightforward to verify that given a finite-dimensional $C^{*}$ algebra $\mathcal{A}$, the set of self-adjoint elements can also be turned into a special JB algebra.
This is a consequence of the fact that finite-dimensional $C^{*}$-algebras can be characterised as direct sums of matrix algebras over the complex numbers \cite{cstaralgebrasbook}. A JB algebra which is isometrically isomorphic to a Jordan subalgebra of a $C^{*}$-algebra is called a JC-algebra. 
    
\end{appendices}

\end{document}